\documentclass[prl,twocolumn,showpacs,amsmath,amssymb,superscriptaddress,shortbibliography]{revtex4-1}
\usepackage[english]{babel}
\usepackage{amsmath}
\usepackage{graphicx, nicefrac}
\usepackage{float}
\usepackage{siunitx}

\usepackage{dcolumn}
\usepackage{bm,graphicx}
\usepackage{etoolbox}

\usepackage[colorlinks]{hyperref}
\usepackage{url}
\usepackage[colorlinks]{hyperref}
\usepackage[normalem]{ulem}
\usepackage[dvipsnames,usenames]{color}
\usepackage{xcolor}
\usepackage[version=3]{mhchem} 

\begin{document}

\title{EuCd$_2$As$_2$: a magnetic semiconductor }

\author{D.~Santos-Cottin}
\email[]{david.santos@unifr.ch}
\affiliation{Department of Physics, University of Fribourg, CH-1700 Fribourg, Switzerland}

\author{I.~Mohelsk\'y}
\affiliation{LNCMI, CNRS-UGA-UPS-INSA, 25, avenue des Martyrs, F-38042 Grenoble, France}

\author{J.~Wyzula}

\author{F.~Le Mardel\'e}
\affiliation{Department of Physics, University of Fribourg, CH-1700 Fribourg, Switzerland}
\affiliation{LNCMI, CNRS-UGA-UPS-INSA, 25, avenue des Martyrs, F-38042 Grenoble, France}

\author{I.~Kapon}
\affiliation{Department of Physics, University of Geneva, CH-1204 Geneva, Switzerland}

\author{S.~Nasrallah}
\affiliation{Department of Physics, University of Fribourg, CH-1700 Fribourg, Switzerland}
\affiliation{Institute of Solid State Physics, TU Wien, A-1040 Vienna, Austria}

\author{N.~Bari\v{s}i\'c} 
\affiliation{Institute of Solid State Physics, TU Wien, A-1040 Vienna, Austria}
\affiliation{Department of Physics, Faculty of Science, University of Zagreb, Bijeni\v{c}ka 32, HR-10000 Zagreb, Croatia}

\author{I.~\v{Z}ivkovi\'c}
\author{J.~R.~Soh}
\affiliation{Institut de Physique, \'Ecole Polytechnique F\'ed\'erale de Lausanne (EPFL), CH-1015 Lausanne, Switzerland}

\author{F.~Guo}
\author{K. Rigaux}
\author{M.~Puppin}
\author{J.~H.~Dil}
\affiliation{Lausanne Centre for Ultrafast Science (LACUS), \'Ecole Polytechnique F\'ed\'erale de Lausanne (EPFL), CH-1015 Lausanne, Switzerland}
\affiliation{Institut de Physique, \'Ecole Polytechnique F\'ed\'erale de Lausanne (EPFL), CH-1015 Lausanne, Switzerland}

\author{B.~Gudac}
\author{Z.~Rukelj}
\author{M.~Novak}
\affiliation{Department of Physics, Faculty of Science, University of Zagreb, Bijeni\v{c}ka 32, HR-10000 Zagreb, Croatia}

\author{A.~B.~Kuzmenko}
\affiliation{Department of Physics, University of Geneva, CH-1204 Geneva, Switzerland}

\author{C.~C.~Homes}
\affiliation{National Synchrotron Light Source II, Brookhaven National Laboratory, Upton,
   New York 11973, USA}

\author{Tomasz~Dietl}
\affiliation{International Research Centre MagTop, Institute of Physics, Polish Academy of Sciences, Aleja Lotnikow 32/46, PL-02668 Warsaw, Poland}
\affiliation{WPI Advanced Institute for Materials Research, Tohoku University, 2-1-1 Katahira, Aoba-ku, Sendai 980-8577, Japan}

\author{M.~Orlita}
\affiliation{LNCMI, CNRS-UGA-UPS-INSA, 25, avenue des Martyrs, F-38042 Grenoble, France}
\affiliation{Institute of Physics, Charles University, CZ-12116 Prague, Czech Republic}
\author{Ana~Akrap}
\email[]{ana.akrap@unifr.ch}
\affiliation{Department of Physics, University of Fribourg, CH-1700 Fribourg, Switzerland}

\date{\today}
\begin{abstract}
EuCd$_2$As$_2$ is now widely accepted as a topological semimetal in which a Weyl phase is induced by an external magnetic field.
We challenge this view through firm experimental evidence using a combination of electronic transport, optical spectroscopy and excited-state photoemission spectroscopy. We show that the EuCd$_2$As$_2$ is in fact a semiconductor with a gap of 0.77~eV. 
We show that the externally applied magnetic field has a profound impact on the electronic band structure of this system. This is manifested by a huge decrease of the observed band gap, as large as 125~meV at 2~T, and consequently, by a giant redshift of  the interband absorption edge. However, the semiconductor nature of the material remains preserved. EuCd$_2$As$_2$ is therefore a magnetic semiconductor rather than a Dirac or Weyl semimetal, as suggested by {\em ab initio} computations carried out within the local spin-density approximation.
\end{abstract}

\maketitle

Magnetic Weyl semimetals have harboured the great hope of bringing spintronics and topology together, 
unfortunately, to a great extent only in theory.
Candidate materials where the magnetic Weyl phase might come to fruition are scarce, and their solid experimental confirmations are even scarcer \cite{Liu:2019,Morali:2019,Nie:2022,Wang:2022,Kanagaraj:2022}.
EuCd$_2$As$_2$ has been seen as one of a few rare magnetic Weyl semimetals -- stable under ambient conditions, with large Eu spins positioned on a frustrated triangular lattice. 
The interplay of frustrated magnetism with topological bands made EuCd$_2$As$_2$ into a hopeful playground for a broad range of exciting phenomena \cite{Ma:2019,Xu:2021}.
Through an extreme sensitivity of valence and conduction bands to Eu magnetism, external fields would then modify the band structure \cite{Soh:2019}. This compound has up to this point been proposed and interpreted as a Weyl semimetal, based upon {\em ab initio} band structure calculations, electronic transport and photoemission measurements \cite{Ma:2019,Soh:2019,Jo:2020,Taddei:2022}.

In this work, we study ultraclean EuCd$_2$As$_2$ crystals.
Electronic transport measurements indicate an extremely low hole concentration.
Accordingly, optical conductivity shows no detectable Drude component, and a strong Reststrahlen phonon mode. 
Our pump-probe photoemission measurement points to a clear band gap, and a carrier lifetime in the picosecond time scale.
These experiments, together with extensive optical measurements, provide decisive proof that EuCd$_2$As$_2$ is a semiconductor, and not a topological semimetal as previously thought \cite{Ma:2019,Ma:2020,Jo:2020,Soh:2018,Soh:2020,Taddei:2022}. We deduce a band gap of 770~meV.
Our results underline the need for more adapted band structure calculations.
Moreover, our results demonstrate that the band structure of EuCd$_2$As$_2$ changes dramatically in an external magnetic field, due to an intimate coupling to localized Eu spins. All of these effects persist up to high temperatures, deep within the paramagnetic phase. 
We understand this through a strong exchange coupling of Eu 4$f$ localized spins to the band states originating from Eu 6$s$, 6$p$ and 5$d$ orbitals.

%
%
\begin{figure*}[!t]
	\includegraphics[width=1\linewidth]{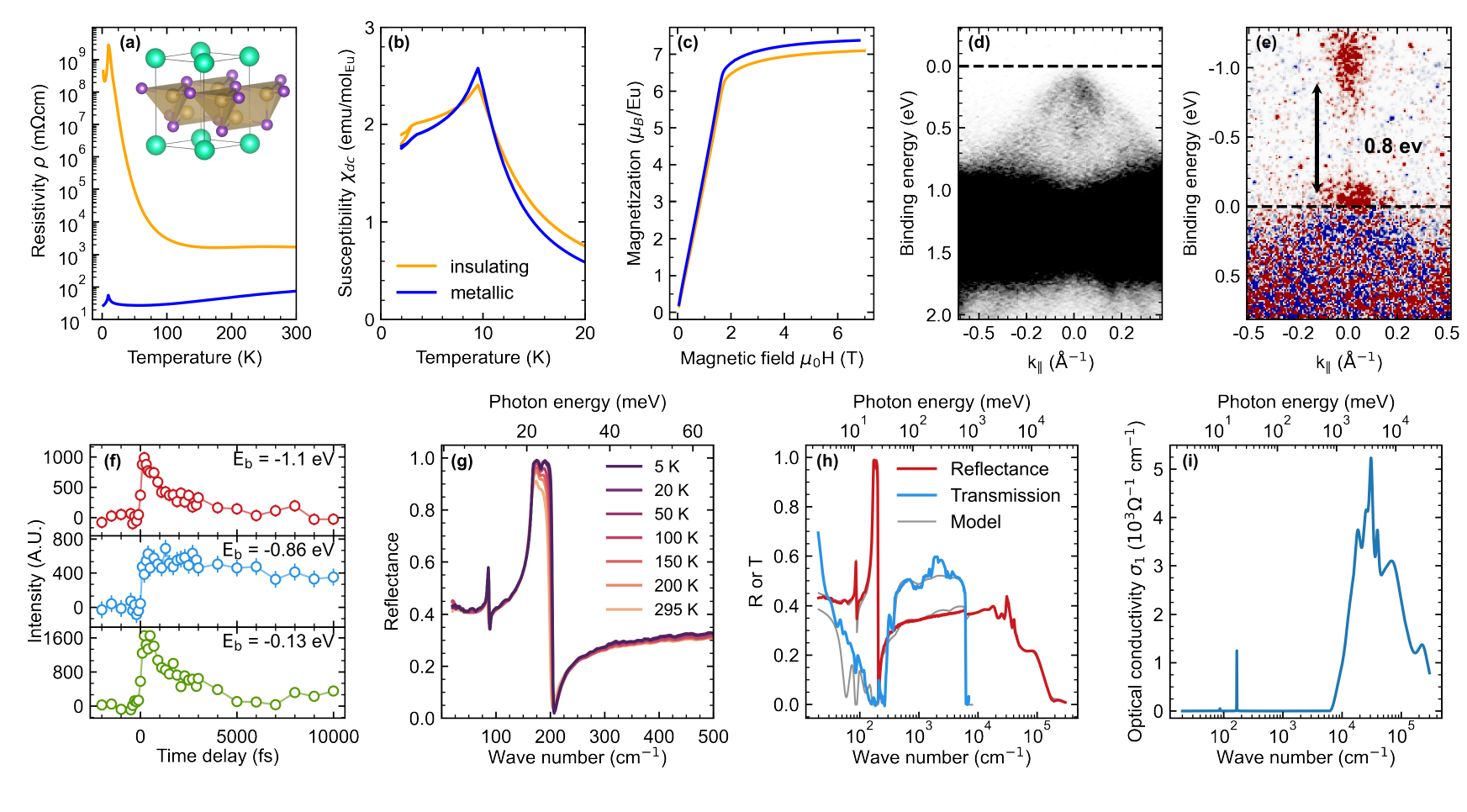}
	\caption{(a) Resistivity as a function of temperature for an insulating and a metallic sample. Inset shows the structure of EuCd$_2$As$_2$, with Eu atoms shown in green, Cd atoms in yellow, and As atoms in purple.
	Magnetic properties, (b) dc susceptibility, and (c) magnetization at 4~K, are shown for the same two sample batches. 
	(d) Static ARPES band map. 
	(e) Pump-probe ARPES experiment results at a delay of 200~fs, where red means increased count rate after the pump-pulse excitation.
	(f) Time traces of the intensity integrated at different energies $E_b$ above the Fermi level. The integration is done for a 200~meV window around the indicated energy.
	(g) Infrared reflectance at different temperatures, for the insulating sample batch.  	
	(h) Transmission and reflectance at 5 K, and their multilayer modelling, which results in (i) the real part of the optical conductivity, $\sigma_1$. 
	}
	\label{fig1}
\end{figure*}

%
%
Single crystals of EuCd$_2$As$_2$ were prepared by the Sn-flux method \cite{Ma:2019}.
Details of sample synthesis, X-ray and electron probe microanalysis are described in Supplemental Material (SM) \cite{SM}.
Carrier density was changed through controlling the purity of the starting materials.
To increase the crystal size and quality, a two-step process was employed, where the crystals from the first growth were used as a seed material for the final growth. 
The layered trigonal lattice of EuCd$_2$As$_2$, shown in the inset of Fig.~\ref{fig1}(a), results in triangular and hexagonally shaped crystals. The flux-growth of single crystals leads to an optically isotropic (001) surface. Blocks of Cd$_2$As$_2$ are sandwiched between the Eu planes, with similar distances to binary semiconducting CdAs$_2$ \cite{Schellenberg:2011}.
We have determined infrared reflectance and transmission \cite{Homes1993,Tanner2015}, complemented by infrared measurements in magnetic fields up to 16~T, for details see SM \cite{SM}.
Time-resolved photoemission (tr-ARPES) experiments were carried out at the EPFL LACUS, at the ASTRA end station \cite{Crepaldi:2017} of the Harmonium beamline \cite{Ojeda2015}, with details described in SM \cite{SM}.
%

%
%
\begin{figure*}[!th]
	\includegraphics[width=1\linewidth]{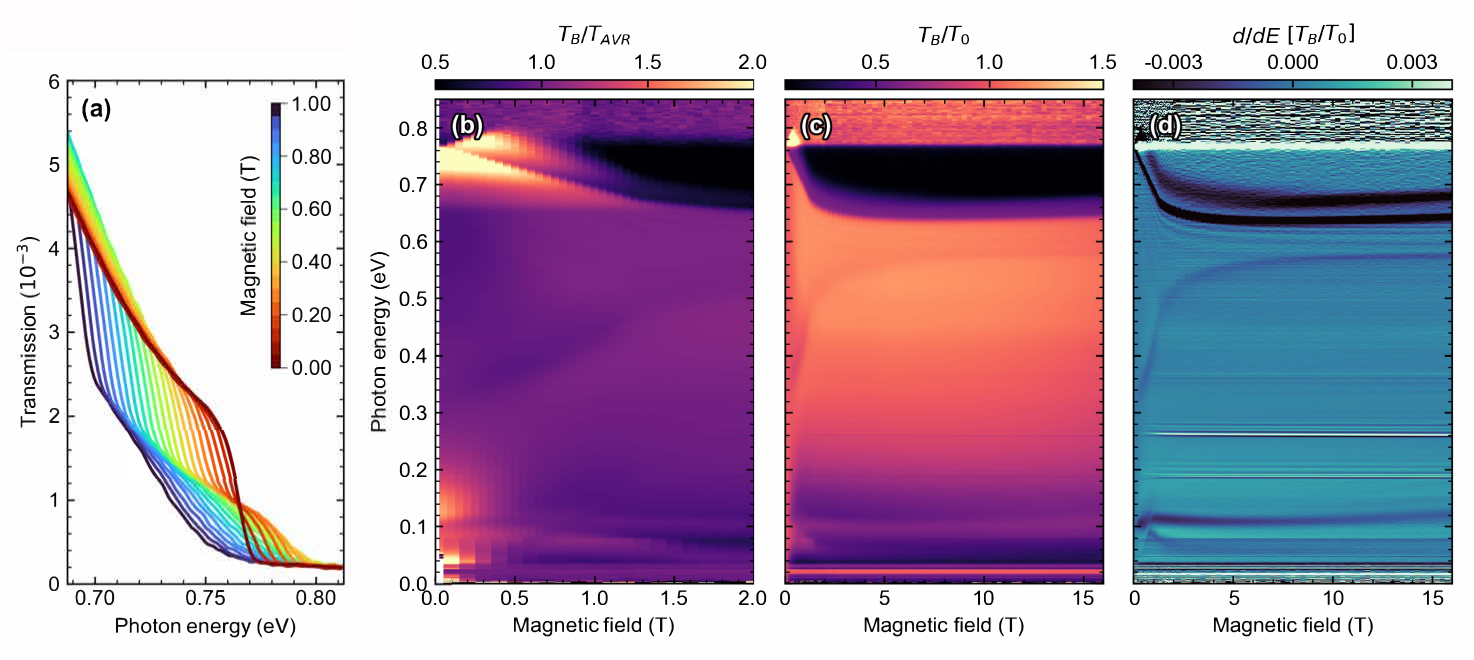}
	\caption{
	(a) Near infrared transmission showing the interband absorption edge at low fields,  $B < 1$~T. 
	(b) Color plot of relative magneto-transmission, $T_B/T_\text{AVR}$, in a broad energy range, and up to 2 T.
	(c) Magneto-transmission $T_B/T_0$ and (d) its first energy-derivative, $d/dE [T_B/T_0]$, in a broad energy and magnetic field range. 
	}
	\label{fig2}
\end{figure*}
Through a controlled crystal synthesis, we obtained insulating samples of EuCd$_2$As$_2$, which have not been reported before. 
Figure~\ref{fig1}(a) shows the resistivity of an insulating sample, resulting from a higher-purity synthesis, compared to a metallic sample, made through a standard purity synthesis. 
The resistivity of a high-purity sample is thermally activated above $T_N$, with an activation energy of $\sim 30$~meV, flattening above 170~K.
Given the Hall coefficient sign, this indicates that thermal activation from the acceptor states is present up to 170~K and valence band transport above. A strong activation energy decrease in a magnetic field, and the corresponding colossal negative magnetoresistance, is expected for magnetic semiconductors in a paramagnetic phase \cite{Jaroszynski:1985,Dietl:2008}, in agreement with our resistivity in magnetic fields \cite{SM}.
The standard synthesis results in metallic behavior above 50~K, consistent with previous reports. 
Interestingly, in both samples the resistivity peak at $T_N=9.5$~K coincides with a sharp, symmetric peak in the magnetic susceptibility, $\chi_{dc}$ in Fig.~\ref{fig1}(b), and an antiferromagnetic (AFM) ordering \cite{Jo:2020}. 
The susceptibility in both metallic and insulating samples shows no difference between the zero-cooled and field-cooled (measured in 10~mT) values above 3.5~K \cite{SM}, excluding a possible ferromagnetic phase above $T_N$ \cite{Artmann1996}. 
The magnetization $M(\mu_0 H)$ is measured with the field applied along the $c$ axis, perpendicular to the Eu planes. 
$M(\mu_0 H)$ first steeply and linearly increases up to 0.8~T, followed by a kink at 1.8~T, reaching a saturated value of $\sim 7\mu_B/$Eu atom. This corresponds to the divalent Eu with half filled 4$f$ orbitals, like in EuTe \cite{Oliveira:1972_PRB}. The initial steep slope of $M(\mu_0 H)$ is consistent with a ferromagnetic structure within each layer, and a relatively weak magnetic anisotropy expected for atoms with a half-filled orbital. 
Despite a strong difference in the resistivity, the susceptibility and magnetization are very similar in both samples, implying a weak effect of residual doping on magnetic properties.
Both samples are insulating or at the localization boundary, meaning there are insufficient band carriers at low temperatures to change $T_N$ by the RKKY mechanism.

The resistivity of the sample with lower residual doping strongly disagrees with the widely accepted notion that EuCd$_2$As$_2$ is a semimetal, and instead suggests it is a semiconductor.
To confirm this, we turn to photoemission measurements of the more resistive samples.
Figure~\ref{fig1}(d) shows a static ARPES band map, where the band structure agrees with the published spectra  \cite{Soh:2019,Ma:2019,Jo:2020}. A strong $4f^7$ band is centered around 1.3~eV below the Fermi level, and the valence band is near the Fermi level.
The results of a pump-probe experiment in Fig.~\ref{fig1}(e) show the conduction band minumum around 800~meV above the Fermi level. No further states are seen in the band gap under these conditions. The conduction band shows extremely high intensity and is visible without any data treatment.
Time traces of the intensity integrated at different energies above the Fermi level are shown in Fig.~\ref{fig1}(f). 
The valence band maximum at $\sim 130$~meV, and the conduction band away from the minimum, both show a time delay of around 1 ps. 
In contrast, the conduction band minimum ($E_b = -0.86$~eV) shows hardly any decay within the full observed time range of 10~ps. The long recombination time implies an energy barrier, and confirms there is a band gap of $770 \pm 70$~meV.
Similar results are obtained on a metallic sample \cite{SM}.

%
%
\begin{figure*}[!htb]
	\includegraphics[width=1\linewidth]{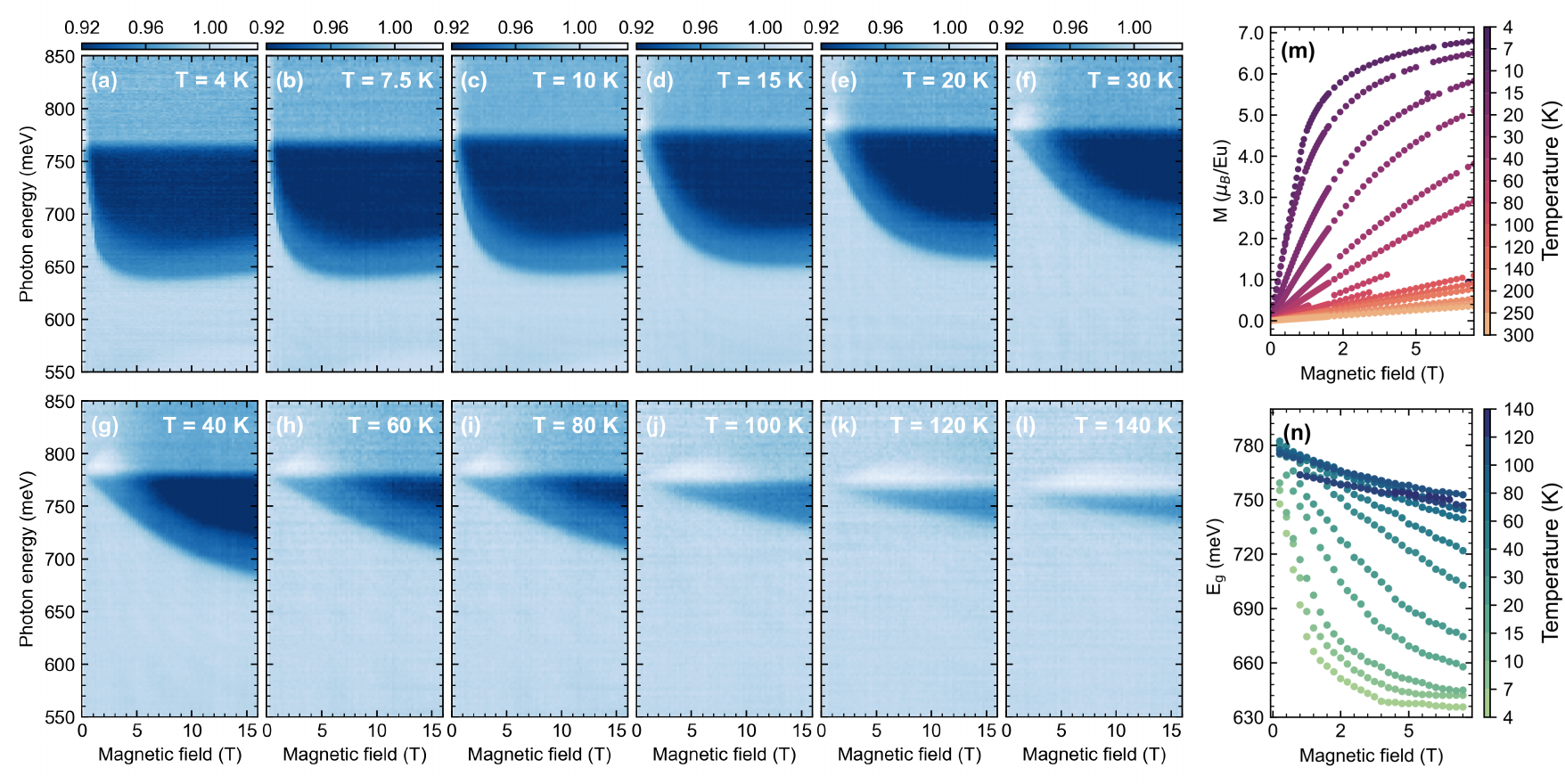}
	\caption{
	(a--l) Colorplots of relative infrared magneto-reflectivity, $R_B/R_0$. Each panel represents a different temperature, going from the AFM phase below 9.5 K, deep into the paramagnetic phase.
	(m) Magnetization as a function of magnetic field, at the same set of temperatures where magneto-reflectivity is shown.
	(n) Onset of absorption extracted from the magneto-reflectivity color plots, at the same set of temperatures as in (m).
	}
	\label{fig3}
\end{figure*}
The infrared properties of EuCd$_2$As$_2$ are shown in Fig.~\ref{fig1}(g-i), determined for the insulating sample. 
Far-infrared reflectance, shown at several different temperatures in Fig.~\ref{fig1}(g), is dominated by two strong in-plane, $E_u$ phonon  modes, at  86 and 165~cm$^{-1}$ \cite{Homes:2023}. The higher frequency mode is an unscreened Reststrahlen mode, since there are no free carriers to screen it. The weak temperature dependence of the reflectance is typical of semiconductors. Moreover, the value of reflectance at low photon energies is far below unity, contrasting the behaviour of metals and semimetals.
To obtain the precise value of the semiconducting band gap, we determine transmission through a thin EuCd$_2$As$_2$ sample.
Transmission and reflectance are modelled using a multilayer model of the dielectric response \cite{Kuzmenko2005}. Kramers-Kronig analysis of reflectance is unreliable as the sample is highly transparent in the mid-infrared range. The obtained optical conductivity, $\sigma_1(\omega)$, is shown in Fig.~\ref{fig1}(i). The onset of absorption coincides with a sudden drop in transmission at 770~meV (6200~cm$^{-1}$). No Drude component appears in the optical conductivity, in line with our resistivity measurement.
Metallic samples \cite{SM} show screened phonons, in agreement with previous optical studies \cite{Wang:2016,Homes:2023}. In metallic and insulating samples, the infrared phonons appear at the same frequencies, and there is a strong increase of the optical conductivity above 1~eV. In both cases, the Drude contribution, if at all present, is minor.
Based on the experimental evidence outlined thus far, we assert that EuCd$_2$As$_2$ is a
semiconductor---in our case with a light $p$-type doping---whose carrier density depends on the starting material purity and the crystal synthesis.
The steep slope of the interband absorption edge at 770~meV is typical of a direct band gap~\cite{CardonaYu}.
This is consistent with the ARPES data in Fig.~\ref{fig1}(e), showing the accumulation of both photoexcited electrons and holes at the $\Gamma$ point.

The electronic band structure of EuCd$_2$As$_2$ is remarkably tunable even with a small magnetic field. 
First, let us focus on the interband absorption edge, evident from magneto-transmission. 
In Fig.~\ref{fig2}a, transmission spectra taken at 4~K are shown at photon energies around the onset, as the magnetic field is increased in 50~mT steps up to 1~T. Relative magneto-transmission in a broader energy and magnetic field range is shown in color plots in Fig.~\ref{fig2}(b-d), where in (d) we show the energy-derivative of the magneto-transmission.  
Stacked plots of relative transmission are shown in the Supplemental Material \cite{SM}.
The sharp zero-field step at 770~meV transforms into a double-step feature already at the lowest fields. 
Below 0.5~T, the lower energy step shifts down in energy, and the higher energy step moves up in energy. 
The steps split by 160~meV/T, giving an effective $g$ factor with an extremely high value of $\sim 1500$ \cite{Kirchschlager:2004}. 
This large splitting originates from exchange interaction rather than Zeeman splitting. 
Above 0.5~T, the higher energy step also begins to redshift. 
The splitting between the two steps saturates at about 50~meV. 
This is a signature of spin polarized bands, which differently absorb light with opposite circular polarization. 
The weak oscillations of the signal below the gap, Fig~\ref{fig2}(d), 
are assigned to giant Faraday rotation induced in our gapped system, with spin-split electronic bands, due to interband absorption of light that differs for opposite circular polarizations \cite{SM,Ohnoutek:2016,Bartholomew:1986}.
Remarkably, the band gap decreases by $\Delta E_{\text{g}} = 125$~meV under 2~T of applied field, reaching a plateau above 5~T. 

While the gap is strongly reduced, it never closes in field, and no band inversion is seen.
Instead, the strong redshift of the band gap  $\Delta E_{\text{g}}(B)$ in Fig~\ref{fig2}(b-d), is proportional to Eu spin magnetization $M(\mu_0 H)$, similarly to EuTe \cite{Schmutz:1978}. 
This means that the molecular field approximation is valid, and $\Delta E_{\text{g}}$ can be written as:
\begin{equation}
\Delta E_{\text{g}} = -\frac{1}{2}{\cal{J}}_{\text{eff}}SM(T,H)/M_S,
\label{eq:DeltaEg}
\end{equation}
where ${\cal{J}}_{\text{eff}}$ is an effective exchange energy between band carriers and Eu spins, $S = 7/2$, and  $M_S$ is saturated Eu spin magnetization. Highly localized $4f$ electrons will weakly hybridize with band states, and ${\cal{J}}_{\text{eff}}$ originates mainly from the intraatomic potential exchange interaction. 
This interaction is ferromagnetic. According to optical spectra of free Eu$^{1+}$ ions in the orbital momentum $L=0$ state (no spin-orbit coupling), ${\cal{J}}_{6s{\text{-}}4f} = 52$, ${\cal{J}}_{6p{\text{-}}4f} = 33$ and ${\cal{J}}_{5d{\text{-}}4f} = 215$~meV \cite{Russell:1941,Dietl:1994}, in accord with chemical trends of $spd$--$f$ exchange energies determined for rare earths in solids \cite{Li:1991}.
Applying the above expression to our data, we obtain  ${\cal{J}}_{\text{eff}} \approx 80$~meV.
Therefore, we propose that the bands undergo a large splitting via $spd-f$ exchange coupling to Eu spins. 
Overall, our results provide a clear indication that the Eu magnetic sublattice controls the electronic band structure of the compound via a strong Eu onsite exchange interaction.

The natural question that follows is whether the observed effects are limited to the low-temperature AFM phase. Figure~\ref{fig3} shows a series of magneto-reflectivity color plots taken up to 16~T, at temperatures ranging from 4 to 140~K.
In the magneto-reflectivity, we see the same kind of features as in the magneto-transmission: the band edge redshifting as magnetic field is applied, and a splitting between the upper and lower band gap edge. 
The zero-field band gap increases first, then decreases with increasing temperature.  
The splitting can be described by an effective $g$ factor, which drops from 1500 at 4~K to 80 at 140~K \cite{SM}.
At all the temperatures up to 100 K, we see that the bands become polarized in high magnetic field, even deep within the paramagnetic phase. 
This shows that EuCd$_2$As$_2$ is a strong paramagnet, and that qualitatively there is no difference in its response whether the zero-field state is an AFM or a paramagnet.
Interestingly, at all temperatures, the extracted band edge as a function of external magnetic field behaves qualitatively very similar to magnetization, see. Fig.~\ref{fig3}(m-n).
This means that the Eq. (\ref{eq:DeltaEg}) remains valid for all temperatures and all magnetic fields. 
%
%

In conclusion, with the full weight of the experimental evidence presented in this Letter, we show decisively that EuCd$_2$As$_2$ is not a topological semimetal, but rather a semiconductor with a band gap of 770~meV. This finding was verified on a number of specimens originating in five different syntheses.
The absorption onset dependence on the magnetic field mimics the shape of magnetization. The band gap strongly decreases in magnetic fields, but it never becomes inverted, and the semiconductor nature of the material remains preserved. 
These results show that more accurate {\em ab initio} studies are desired, scrutinizing in detail all the complexities of correlated electron physics in Eu-based compounds \cite{Cuono:2023}.
Nonetheless, the local Eu magnetic moments are responsible for band structure changes, through strong 4$f$ exchange coupling to valence and conduction states.

%
%
We thank B. Fauqu\'e, M. M\"uller, D. van der Marel, N. Schr\"oter, A. Grushin, F. de san Juan,  Iurii Timrov and Luca Binci for enriching conversations.
We are grateful to N. Miller for the helpful comments.
A.A. acknowledges funding from the  Swiss National Science Foundation through project PP00P2\_202661.
This research was supported by the NCCR MARVEL, a National Centre of Competence in Research, funded by the Swiss National Science Foundation (grant number 205602).
F.G. and J.H.D. acknowledge the funding through SNSF grant 200021\_200362.
M.N. and N.B. acknowledge the support of CeNIKS project co-financed by the Croatian Government and the EU through the European Regional Development Fund Competitiveness and Cohesion Operational Program (Grant No. KK.01.1.1.02.0013).
This work has been supported by the ANR DIRAC3D. We acknowledge the support of LNCMI-CNRS, a member of the European Magnetic Field Laboratory (EMFL).
This work was supported by the Foundation for Polish Science through the International Research
Agendas program co-financed by the European Union within the Smart Growth Operational Programme.
The work at the TU Wien was supported by the European Research Council 
(ERC Consolidator Grant No 725521).

\bibliography{EuCd2As2}

\newpage

\newpage
\vspace*{-2.0cm}
\hspace*{-2.5cm} {
  \centering
  \includegraphics[width=1.2\textwidth]{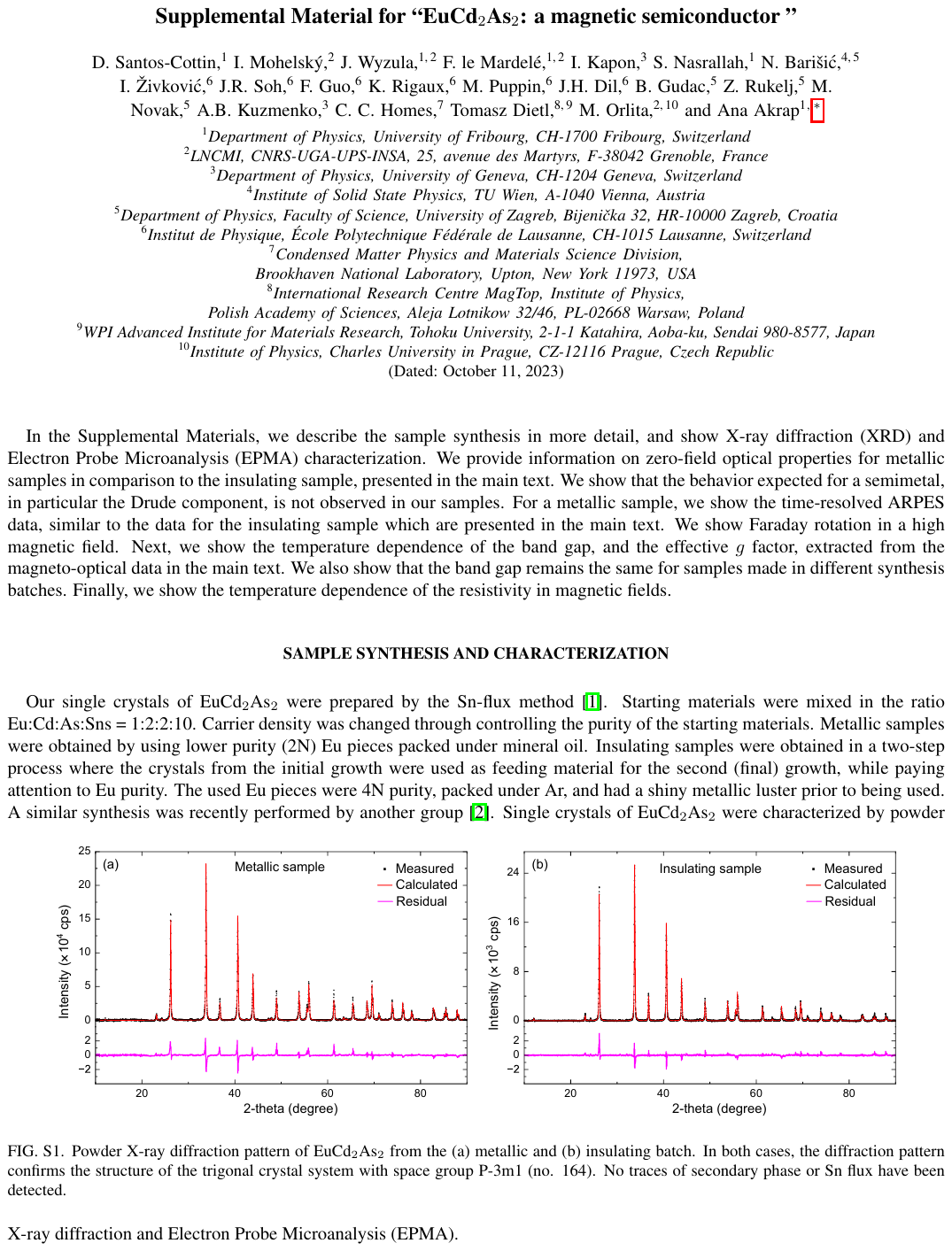} \\
  \ \\
}

\newpage
\vspace*{-2.0cm}
\hspace*{-2.5cm} {
  \centering
  \includegraphics[width=1.2\textwidth]{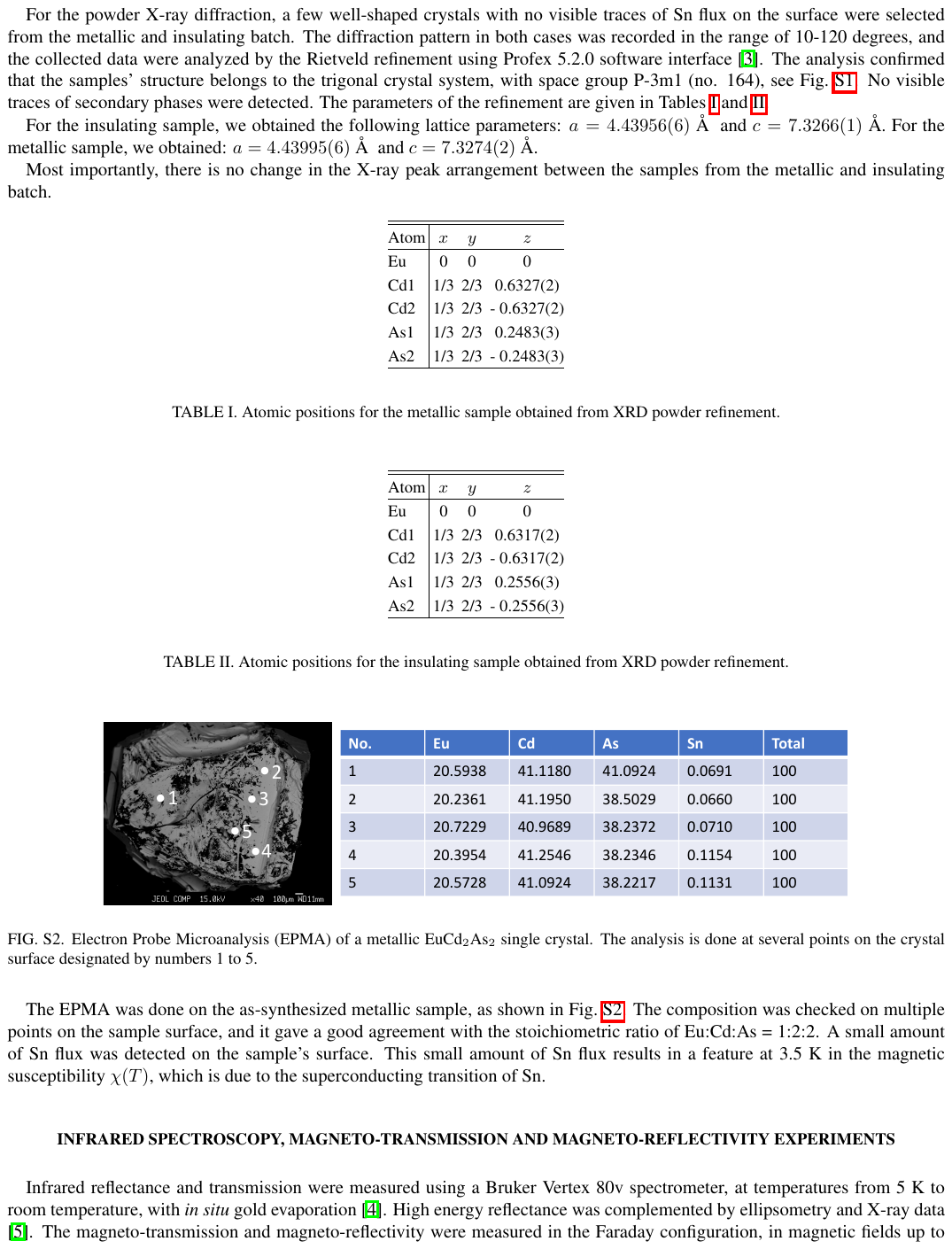} \\
  \ \\
}

\newpage
\vspace*{-2.0cm}
\hspace*{-2.5cm} {
  \centering
  \includegraphics[width=1.2\textwidth]{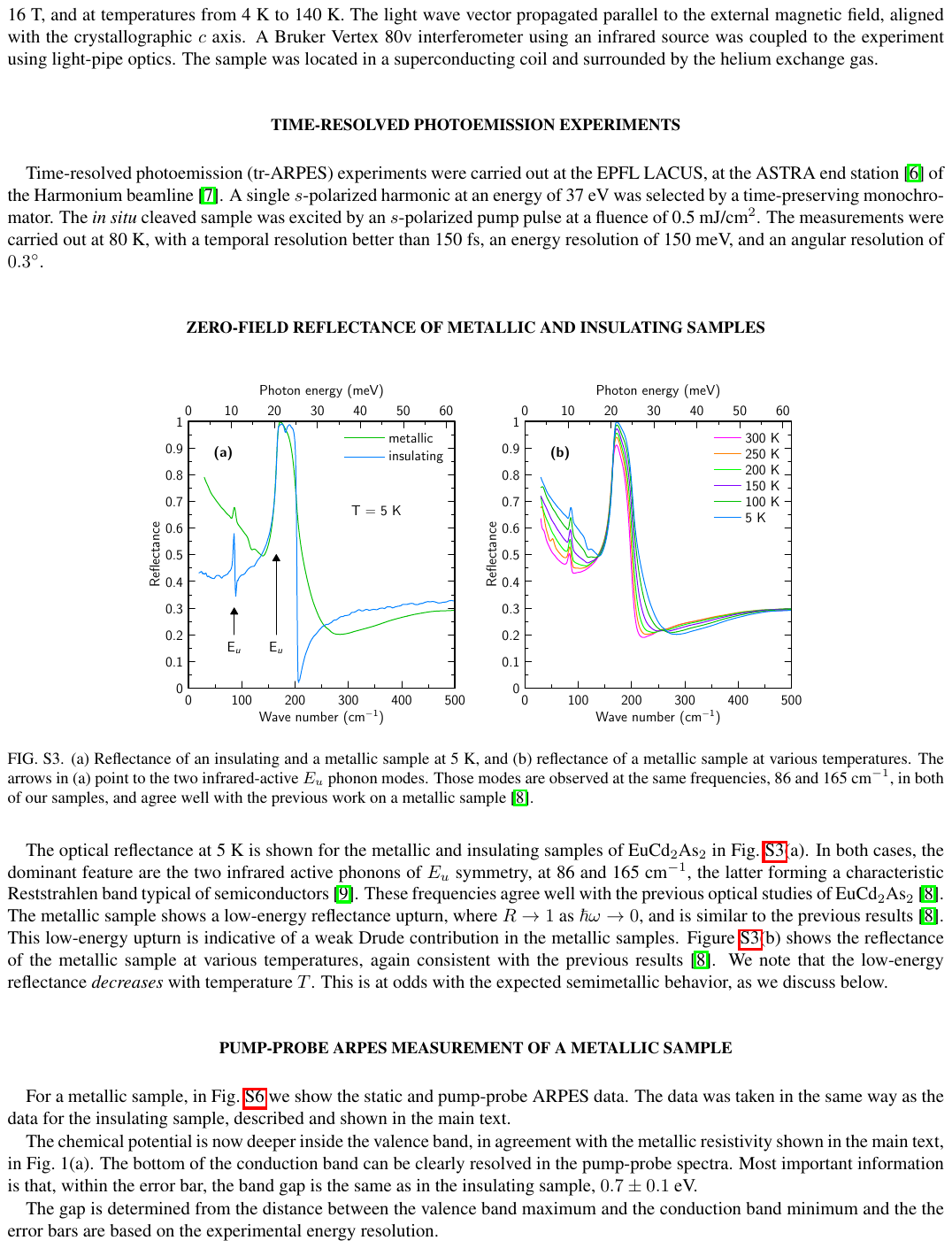} \\
  \ \\
}

\newpage
\vspace*{-2.0cm}
\hspace*{-2.5cm} {
  \centering
  \includegraphics[width=1.2\textwidth]{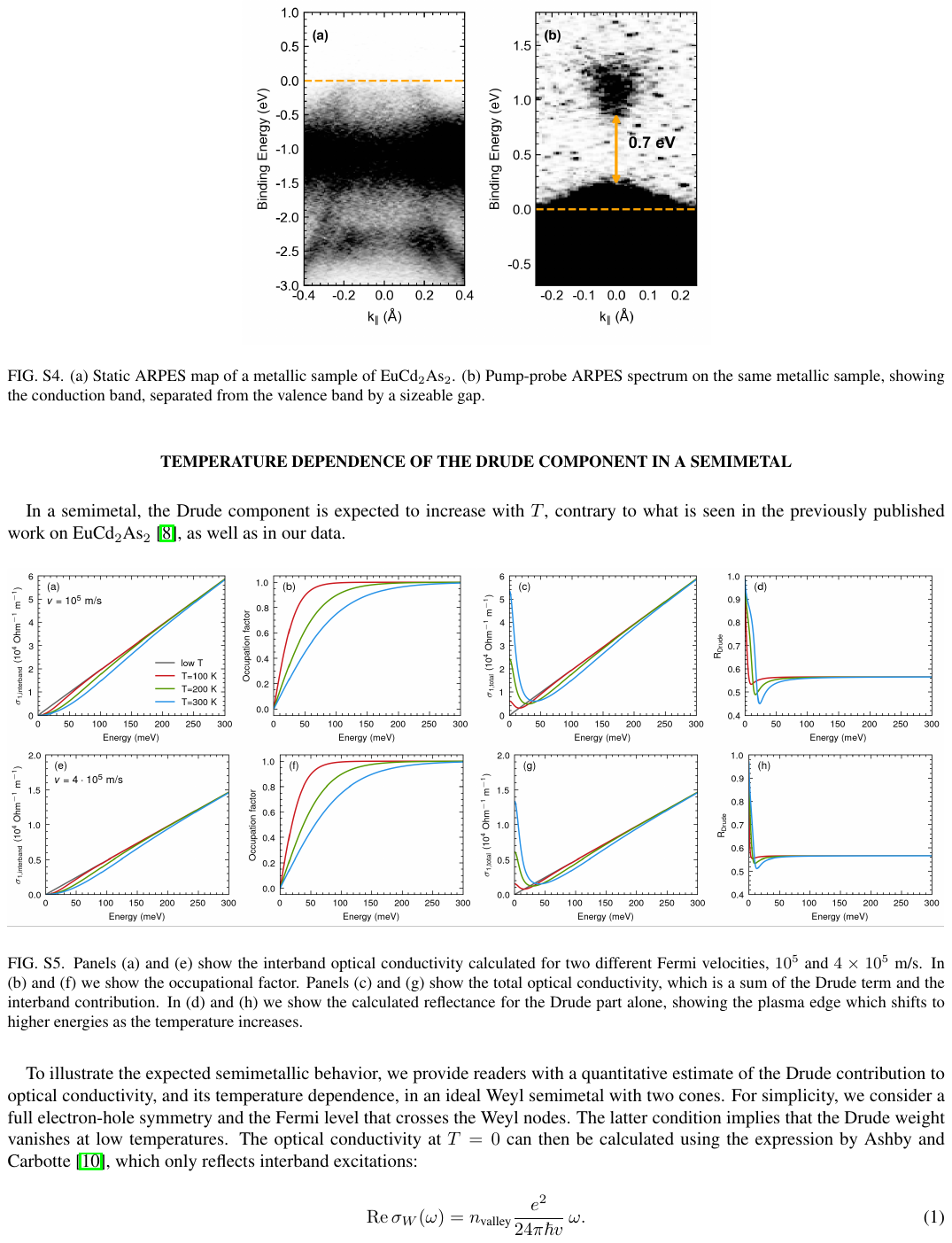} \\
  \ \\
}

\newpage
\vspace*{-2.0cm}
\hspace*{-2.5cm} {
  \centering
  \includegraphics[width=1.2\textwidth]{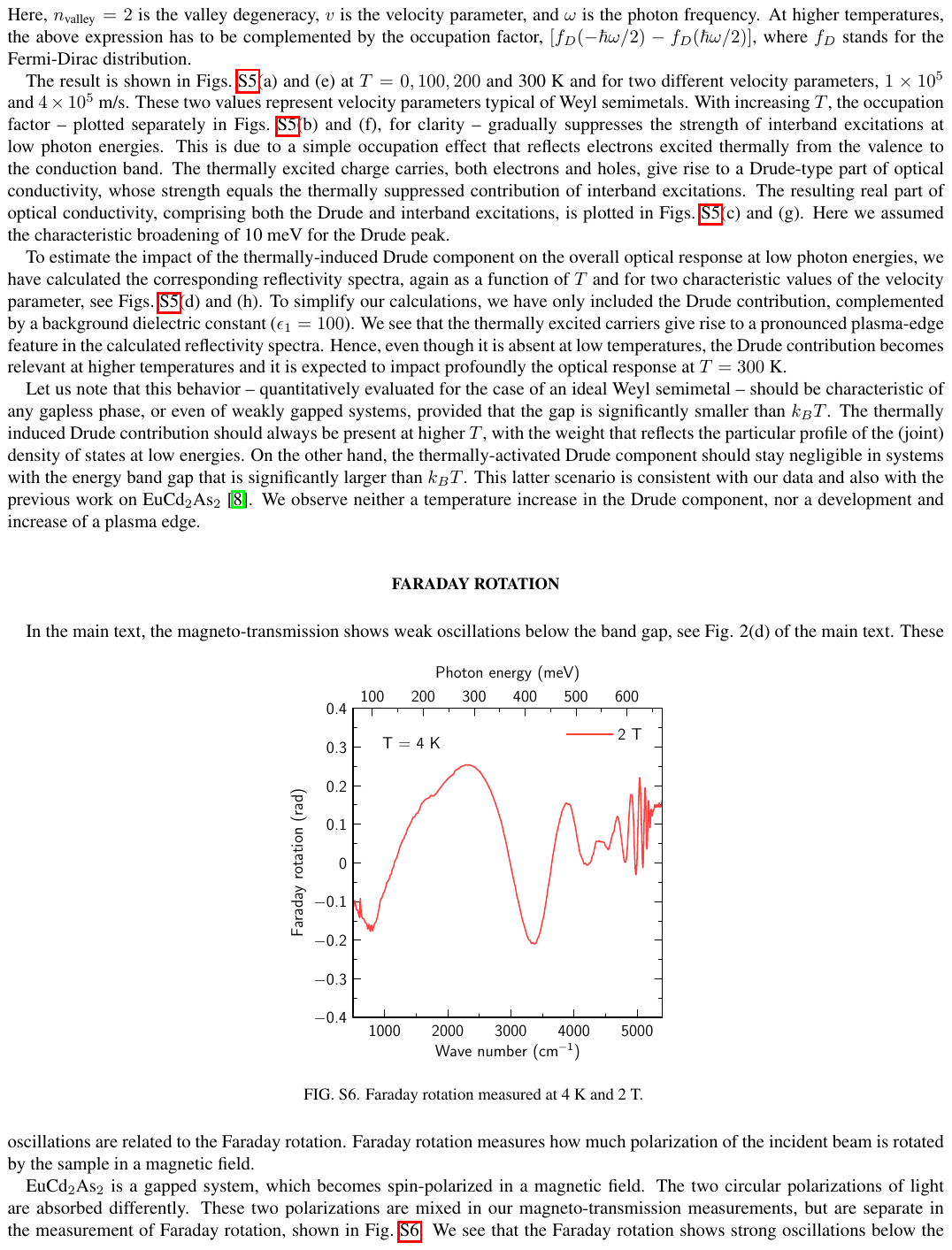} \\
  \ \\
}

\newpage
\vspace*{-2.0cm}
\hspace*{-2.5cm} {
  \centering
  \includegraphics[width=1.2\textwidth]{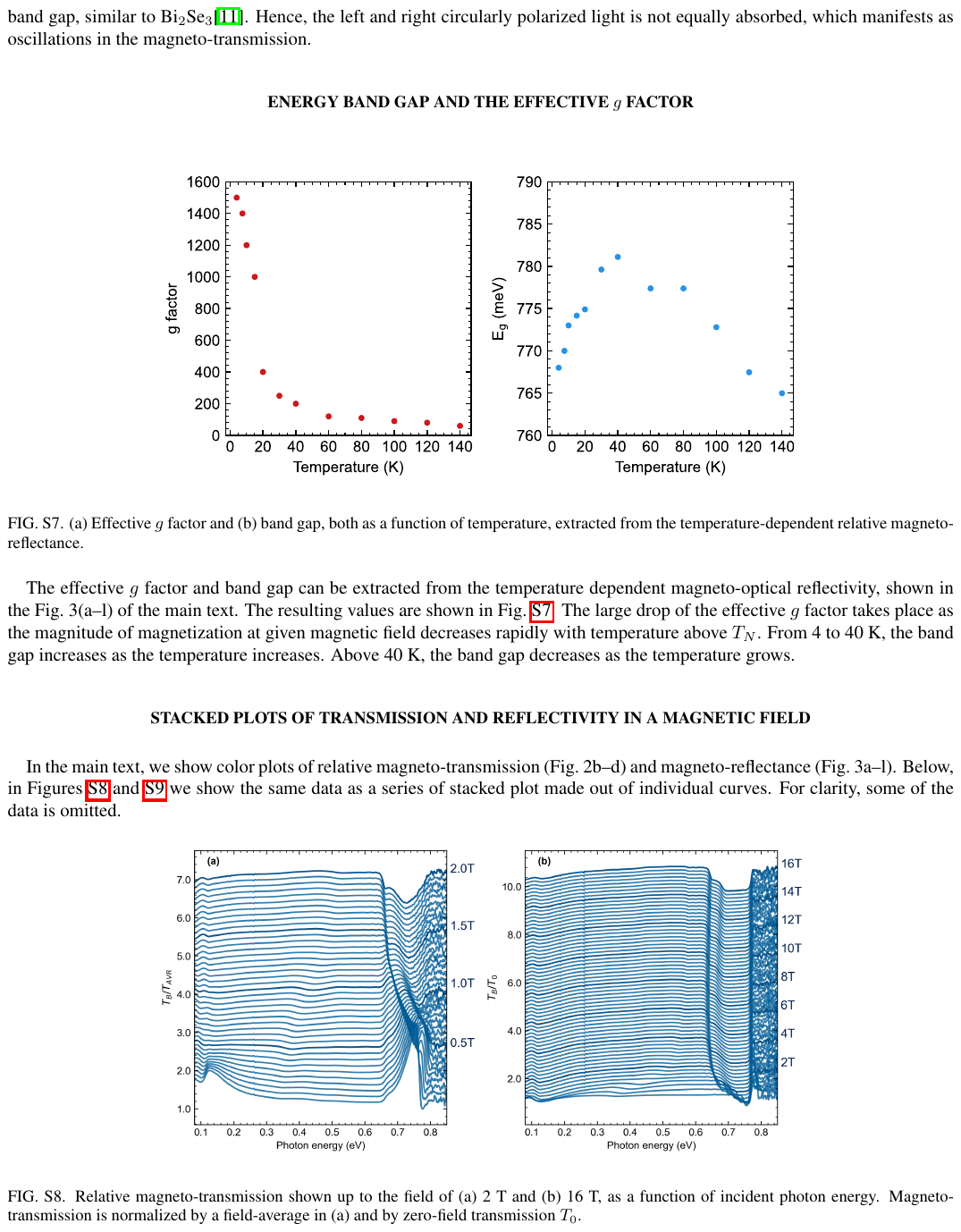} \\
  \ \\
}

\newpage
\vspace*{-2.0cm}
\hspace*{-2.5cm} {
  \centering
  \includegraphics[width=1.2\textwidth]{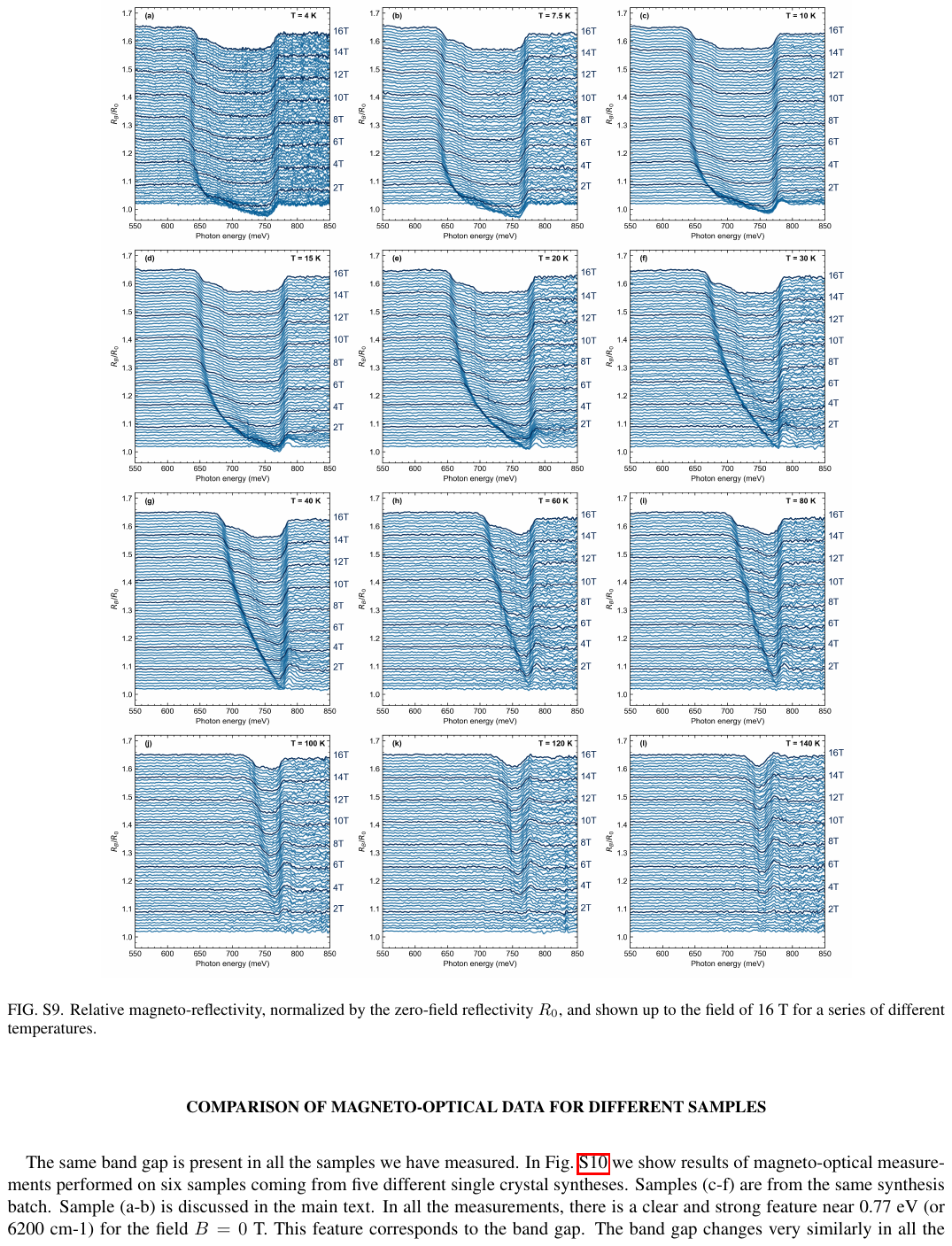} \\
  \ \\
}

\newpage
\vspace*{-2.0cm}
\hspace*{-2.5cm} {
  \centering
  \includegraphics[width=1.2\textwidth]{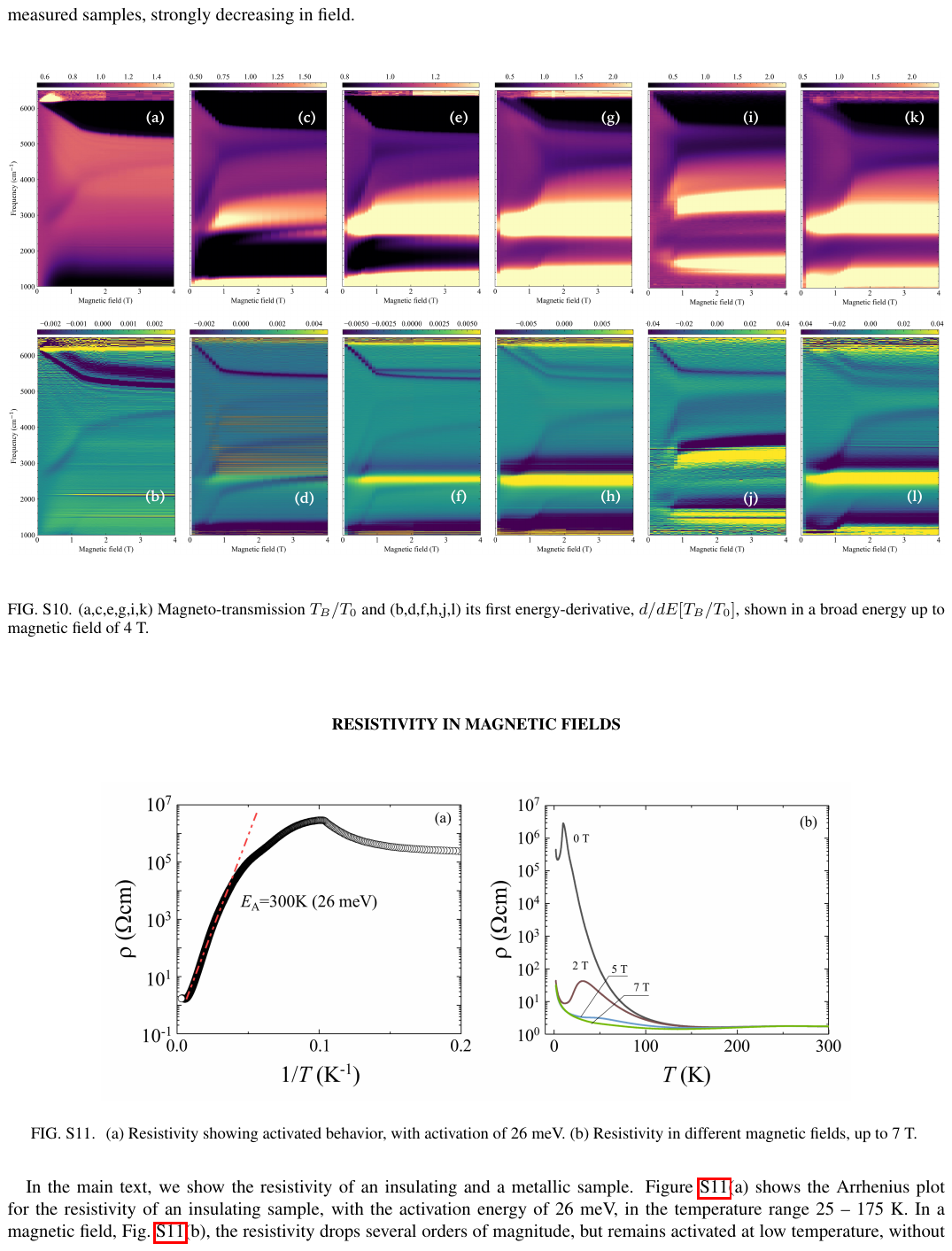} \\
  \ \\
}

\newpage
\vspace*{-2.0cm}
\hspace*{-2.5cm} {
  \centering
  \includegraphics[width=1.2\textwidth]{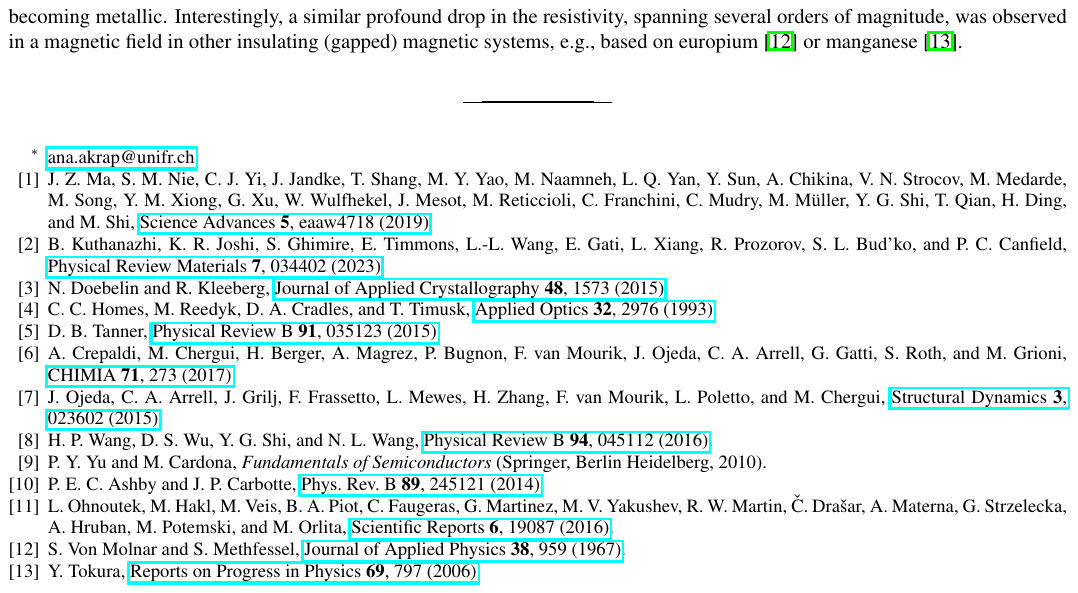} \\
  \ \\
}

\end{document}